\documentclass{vgtc}                          

\ifpdf
  \pdfoutput=1\relax                   
  \usepackage{graphicx}                
  \DeclareGraphicsExtensions{.pdf,.png,.jpg,.jpeg} 
\else
  \usepackage{graphicx}                
  \DeclareGraphicsExtensions{.eps}     
\fi%

\graphicspath{{figures/}{pictures/}{images/}{./}} 

\usepackage{microtype}                 
\PassOptionsToPackage{warn}{textcomp}  
\usepackage{textcomp}                  
\usepackage{mathptmx}                  
\usepackage{times}                     
\usepackage{cite}                      
\usepackage{tabu}                      
\usepackage{booktabs}                  

\usepackage{tabularx}
\usepackage{amsmath}
\usepackage{amssymb}
\usepackage{hyperref}

\DeclareMathOperator*{\argmin}{arg\,min}
\newcommand{\clr}{\textcolor{black}}

\onlineid{0}

\vgtccategory{Research}

\vgtcinsertpkg

\title{Efficient Interpolation-based Pathline Tracing with B-spline Curves in Particle Dataset}

\author{Haoyu Li \thanks{e-mail: li.8460@osu.edu}\\ %
    \scriptsize The Ohio State University %
\and Tianyu Xiong \thanks{e-mail: xiong.336@osu.edu} \\ %
    \scriptsize The Ohio State University %
\and Han-Wei Shen \thanks{e-mail: shen.94@osu.edu} \\ %
    \scriptsize The Ohio State University %
}

\abstract{Particle tracing through numerical integration is a well-known approach to generating pathlines for visualization. However, for particle simulations, the computation of pathlines is expensive, since the interpolation method is complicated due to the lack of connectivity information. Previous studies utilize the $k$-d tree to reduce the time for neighborhood search. However, the efficiency is still limited by the number of tracing time steps. Therefore, we propose a novel interpolation-based particle tracing method that first represents particle data as B-spline curves and interpolates B-spline control points to reduce the number of interpolation time steps. We demonstrate our approach achieves good tracing accuracy with much less computation time.%
} 

\CCScatlist{
  \CCScatTwelve{Human-centered computing}{Visualization}{Visualization application domains}{Scientific visualization}
}

\begin{document}


\firstsection{Introduction}

\maketitle

Pathline tracing is a popular technique for flow visualization and analysis,  for its ability to depict scientific features in different domains such as aerodynamics and cosmology. 
\clr{Eulerian and Lagrangian flow simulations generate flow fields in two different formats: mesh-based and particle-based. For the meshed-based flow fields, velocities are stored at the vertices of the mesh, while the particle-based method stores the velocity information at particle positions without explicit connectivity defined between particles. }
For certain particle simulations such as smoothed particle hydrodynamics (SPH), even though the trajectories of the particles already exist, to have a clear view of the underlying features, sometimes it is necessary to have more pathlines around the locations of interest.
The widely used numerical integration approaches, such as Euler or Runge-Kutta methods for computing pathlines, require an extensive amount of velocity interpolation both spatially and temporally to obtain reasonable accuracy. The computation time of numerical integration largely depends on the number of integration steps and the complexity of the interpolation method. For large-scale datasets, the I/O overhead can also be very high.  
Therefore, for particle-based flow fields, using numerical integration to generate pathlines is very expensive.

However, the positions of a particle, or the particle displacement, over time often show particular patterns governed by the underlying physical conditions. 
Making use of the particle displacement, previous study \cite{Chandler2015} proposed an interpolation-based tracing approach for particle datasets, where they avoid performing velocity interpolations at every time step needed by numerical integration. Instead, they directly use particle displacement between time steps to generate particle traces. 
The performance of their method is determined by three factors: neighborhood search time, the number of interpolated pathlines, and the number of interpolation steps. 
Neighborhood search time is optimized using a modified $k$-d tree data structure in their work and pathline advection from different starting points can be parallelized. 
However, the tracing time for each pathline cannot be easily reduced. 
Considering that current flow simulations can output data at higher spatial and temporal resolutions, a large number of tracing steps introduces a computational challenge.

To solve the aforementioned challenge, we propose a new approach for interpolation-based pathline tracing. B-spline\cite{gordon1974b} curves are used first to fit the traces of existing particle data.
We optimize the accuracy by using an adaptive knot placement method \cite{yeh2020fast} in B-spline approximation, where more control points are placed at the positions of higher complexity in the curve. 
Interpolation is performed only between the control points of the parametric curves instead of the original particles, which reduces the number of advection iterations from the number of particle integration steps to the number of control points used. 
With our method, we demonstrate a significant reduction in the computation cost for particle tracing while achieving similar tracing quality.

\section{Related Works}

Most of the approaches to calculating pathlines rely on particle tracing by numerical integration \cite{mcloughlin2010over, post2002feature}.
Since regular grid flow field particle tracing is a well-studied area with mature techniques, while studies on particle-based flow field pathline generation are relatively scarce, we focus on the related works to pathline generation for particle-based flow fields in this section. 
One of the most relevant studies to our work is interpolation-based pathline tracing\cite{Chandler2015}, where next time step particle location is calculated based on the neighborhood particle position displacement. Their method avoids multiple neighborhood searches needed by numerical integration at each time step. However, their approach faces the problem that the total tracing time is dominated by the number of time steps of the existing pathlines. 
Particle-based flow fields can also be interpreted as flow maps stored at scattered point locations. Many recent research that use machine learning techniques to generate \cite{han2021exploratory, lee2021deep} or super-resolve flow maps \cite{jakob2020fluid,agus_euro} are also related to this study. However, machine-learning-based methods usually face the problem of long training time and unbounded tracing errors.

Another related field to our study is using parameter curves to represent pathlines or streamlines in flow visualization. An exploratory research \cite{bujack2015lagrangian} examine different kind of parameter curves and their fitting quality. Hong et al. \cite{hong2017compression} make use of B\'ezier curves for flow visualization under a compression-based pathline or streamline reuse framework. Liu and Wang \cite{liu2022topological} proposed a streamline compression method using B-spline curves that preserves topological relations and has bounded error. Chen et al\cite{chen2015uncertainty} model pathlines with composite B\'ezier curves with uncertainty and reduce error using forward and backward trace.
In all of these studies, parameter curves show small and controllable fitting errors in representing pathlines and streamlines and fast fitting time.



\section{Background}
\label{sect:background}
In this section, we review the interpolation-based pathline tracing method introduced by Chandler et al. \cite{Chandler2015} for particle-based flow fields. Each pathline in the dataset can be described as a function: $f_i: \tau \mapsto \rho$,
where $\tau\in\mathbb{N}$ is the time step and $\rho\in\mathbb{R}^3$ is the particle position at that time step. 
When a new particle is inserted at time $\tau$, to compute its trajectory in other time steps, an interpolation-based pathline tracing method  described in the following steps is used:
\begin{enumerate}
  \setlength\itemsep{-0.3em}
  \item Load all particles of time step $\tau$.
  \item Find neighbor particles around the inserted particle.
  \item Load all particles of time step $\tau+1$.
  \item Calculate the interpolation weights based on the spatial positions of the neighboring particles and the inserted particle at time $\tau$.
  \item Reconstruct the position of inserted pathline at time $\tau + 1$ based on neighboring particles' displacement.
  \item Update the neighbor particles at time step $\tau+1$.
  \item Repeat from step 3 until the desired time step is reached.
\end{enumerate}
The reconstruction step in $5$ can be expressed by the following equation:
\begin{equation}
    \label{equ:reconstruction}
    f_{insert}(\tau+1) = \sum_{f_i(\tau)\in N(f_{insert}(\tau))}w_i(f_i(\tau+1)-f_i(\tau))+f_{insert}(\tau),
\end{equation}
where we use $N(f_{insert}(\tau))$ to denote the set of neighboring particles of the particle at position $\rho=f_{insert}(\tau)$ and $w_i$ to denote the interpolation weight.
It is worth noting that the interpolation-based pathline tracing approach is agnostic to the interpolation method and $w_i$ is calculated using the interpolation method of choice, for example, SPH kernels in their study. 
A modified $k$-d tree data structure is constructed at every time step to accelerate neighborhood search in their proposed method. 

The major limitation of this method is that the total computation time is determined by the number of integration (time) steps needed to calculate the pathline. 
Our method aims to solve this limitation by applying interpolation-based pathline tracing on B-spline representations of the existing particle traces. The interpolation iteration needed is effectively reduced to the number of control points used to represent the particle traces.

\section{Method} 

Our B-spline curve interpolation-based pathline tracing approach can be described at a high level in two steps. First, we process the particle-based flow field and fit a B-spline curve for each existing particle trace. Second, we perform interpolation-based pathline tracing on the control points and knots for a given new particle position and time. We explain these two steps in detail next.

\subsection{B-spline Approximation for Particle Traces}

A B-spline curve of order $k$ is a piecewise polynomial function of degree $k-1$ defined by:
\begin{equation}
    C(u) = \sum^{n-1}_{i=0}B_{i,k}(u)P_i, \;\;\;\;\; u \in [t_0,t_{n+k-1}],
\end{equation}
where $P_i$ denotes one of the $n$ control points and $C(u)$ evaluates the B-spline curve at parameter location $u$. Knot vector $T=\{t_0,t_1,t_2,...,t_{n+k-1}\}$ defines the parameter $u$ range that is influenced by the control points $P$. The B-spline basis function $B_{i,k}(u)$ is defined recursively with respect to the knot vector and the curve order $k$ as:
\begin{equation}
\begin{aligned}
    B_{i,1}(u) = &\begin{cases}
    1, & \text{if $t_{i} \leq u < t_{i+1}$} \\
    0, & \text{otherwise} \\
    \end{cases},\\
    B_{i,k}(u) = &\frac{u-t_i}{t_{i+k-1}-t_i}B_{i,k-1}(u) + \frac{t_{i+k}-u}{t_{i+k}-t_{i+1}}B_{i+1,k-1}(u).
\end{aligned}
\end{equation}
More in-depth content about B-spline curves can be found in this book by Farin\cite{farin2002curves}.

Given a pathline represented by $m$ particles whose positions are $\rho_0, \rho_1, ..., \rho_{m-1}$, the approximation of a B-spline curve involves three steps: First, we need to parameterize the data points into a monotonically increasing list $u_0, u_1,...,u_{m-1}$, which determines the distribution of data points along the B-spline curve. 
Second, we determine the knot vector along the curve, which decides the control points' distribution in the parameter space.
And last, we optimize the control point positions with respect to the parameters and the knot vector as a least-square solution:
\begin{equation}
    \label{equ:least_square}
    \argmin_P\sum_0^{m-1}{\lVert \rho_i - C(u_i) \rVert}_2
\end{equation}

\subsubsection{Parameterization}
There are two ways to parameterize this trajectory. The first is to use the time steps as parameters: $\tau$ as the parameter for $\rho$. 
And the second is to use the chord length in 4D space-time to determine a parameter for each particle.
\clr{We choose to use time steps to parameterize the trajectory for its simplicity. When representing the curve in 4D space, we need to determine how to normalize the spatial dimensions and the temporal dimension. The choice of different normalizations may have an impact on the fitting accuracy. Moreover, parameterizing the pathlines by the 4D chord length also requires us to use 4D control points to describe the B-spline curve, which can be less efficient when applying the interpolation algorithm.}
Therefore, we normalize the time index to be in the range $[0,1]$, and each B-spline curve is a function from time to 3D spatial positions.

\subsubsection{Knot Placement}

Considering a B-spline curve of order $k$, control point $P_i$ is used when calculating the spline segment between $t_i$ and $t_{i+k}$. Thus, knot vector $T=\{t_0,t_1,t_2,...,t_{n+k-1}\}$ determines the control point density along the pathline in the parameter space. We duplicate the first $k$ and the last $k$ knots to ensure the curve passes through the first and the last control points. 
Knot placement is a well-known problem in B-spline approximation to optimize the spline quality. We adopted a recent fast automatic knot placement method proposed by Yeh et al.\cite{yeh2020fast} to determine an optimal knot vector for spline fitting. 

The automatic knot placement method is inspired by the idea that an order-$k$ B-spline curve has a piecewise constant $(k-1)$th derivative\cite{yeh2020fast}, and derivative discontinuities mark the knot locations. A feature function is derived from the $k$th derivative of the data points and used the cumulative feature function to guide the knot placement (more knots when the cumulative feature function changes fast). Their method is empirically found to yield better fitting accuracy than directly using $(k-1)$th derivative of the data points. 

The only hyper-parameter to choose in automatic knot placement is the number of knots used. Choosing the number of knots also determines the number of control points used for B-spline approximation, since the number of knots is $n+k$ for a B-spline with $n$ control points and order $k$. This choice is a balance between pathline tracing efficiency and accuracy. We discuss in detail the evaluation of the choice in \autoref{sect:fitting_eval}.
After a knot vector has been calculated for the given data points, we find the control points of the approximated B-spline curve using the least square method following \autoref{equ:least_square}. 

\subsection{Interpolation-Based B-spline Tracing}

\begin{figure}
 \centering
 \vspace{-30pt}
 \includegraphics[width=\columnwidth]{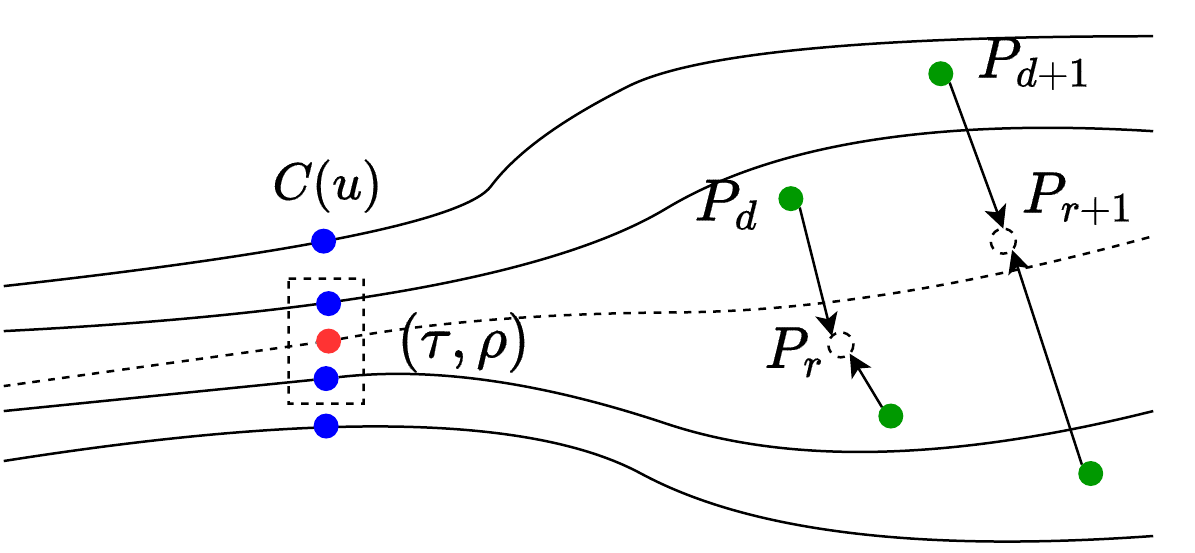}
 \caption{
    An example for forward tracing of control points.
    Given the pathline tracing starting point (red dot), we first identify the neighbor curves by evaluating $C(u)$ (blue). 
    We find the nearest knots and control points (green) to $C(u)$ in future time steps.
    And then neighbor particles (blue dots in the square) and the nearest knots and control points $P_{d}$ are used to reconstruct knots and control points $P_{r}$ for the new pathline. 
    We then similarly interpolate knots and control points to get the next iteration knots and control points $P_{r+1}$ for the new pathline.
 }
 \label{fig:b-spline_interp}
\end{figure}

Based on the method introduced by Chandler et al.\cite{Chandler2015}, we propose a pathline tracing approach based on B-spline control point interpolation. All particle data are processed first and a B-spline curve $C$ is generated for each particle trajectory.
The B-spline approximation only needs to be done once. 
The control points $P_0, P_1,..., P_{n-1}$ and the knot vector $T: t_0, t_1,..., t_{n+k-1}$ are saved as the representation for the data instead of the original particle trajectories. 
\clr{Since the automatic knot placement algorithm cannot guarantee that the knots are placed at the same time steps for different pathlines, we need to synchronize the control points before interpolation. However, for higher efficiency, this synchronization is only performed at the starting point of the traced pathline. For later control points, we simply choose the nearest neighbors among the next control point from each pathline, even though these control points may correspond to the knots of different time steps. The errors introduced by this choice are small for two reasons. First, nearby pathlines will likely share similar shapes so that the knot intervals of nearby pathlines are similar. 
Second, we interpolate both the knots and the control points, which is similar to interpolating a point in 4D space-time, which eases the problem of unsynchronized time steps. }

Next, we describe our algorithm in detail.
An illustration of the interpolation scheme can be found in \autoref{fig:b-spline_interp}.
Given a tracing starting point $(\tau, \rho)$, we present the modified interpolation-based pathline tracing as follows:
\begin{enumerate}
  \setlength\itemsep{-0.3em}
  \item Normalize $\tau$ to get the parameter $u$ and evaluate all existing B-spline at $u$.
  \item For each existing B-spline, find the minimum index $d\in [\lfloor k/2 \rfloor, n +k -1 - \lceil k/2 \rceil]$ so that $t_d \ge u$. For special cases, where $u = 0$ and $1$, we define $d=\lfloor k/2 \rfloor$ and $n +k-1 - \lceil k/2 \rceil$.
  \item Find neighboring B-splines $C$ and calculate the interpolation weights based on ${\lVert \rho - C(u)\rVert}_2$.
  \item Reconstruct the knot $t_r$ and the control point $P_{r-\lfloor k/2 \rfloor}$ of $Q$ based on neighbor $t_d$ and $P_{d-\lfloor k/2 \rfloor}$, where $r$ denotes the knot and control point indices for the interpolated pathlines.
  \item Load knots $t_{d+1}$ and control points $P_{d-\lfloor k/2 \rfloor + 1}$.
  \item Find the neighbor control points and knots. Calculate interpolation weights for iteration $r$.
  \item Reconstruct the knot $t_{r+1}$ and the control point $P_{r-\lfloor k/2 \rfloor+1}$ for the inserted pathline based on the interpolation weights.
  \item Update the neighbor knots and control points for iteration $r+1$.
  \item Repeat from step 5 until $t_r > 1$. 
\end{enumerate}

The steps above describe how we perform forward tracing. Backward tracing, which is necessary to represent the full new B-spline, can be described similarly. The reconstructions in step 4 and step 7 are defined similarly to \autoref{equ:reconstruction}:
\begin{equation}
\begin{aligned}
    \label{equ:reconstruction_2}
    t_{r} &= \sum_i w_i \cdot t_{i,d_i}, \\
    P_{r-\lfloor k/2 \rfloor} &= \sum_i w_i(P_{i,d_i-\lfloor k/2 \rfloor}-C_i(u))+\rho, \\
    t_{r+1} &= \sum_{i}w_i(t_{i,d_i+1}-t_{i,d_i})+t_{r}, \\
    P_{r-\lfloor k/2 \rfloor+1} &= \sum_i w_i (P_{i,d_i-\lfloor k/2 \rfloor+1}-P_{i,d_i-\lfloor k/2 \rfloor})+P_{r-\lfloor k/2 \rfloor}, \\
\end{aligned}
\end{equation}

where we use $i$ to denote the indices of all neighbors of the inserted pathline. In the first two equations, neighbors are calculated based on the Euclidean distance between the positions of the synchronized first time step. In the last two equations, neighbors are calculated using Euclidean distance between the control points of different B-spline curves.
Similar to the original interpolation-based method, we can use any interpolation method to calculate the interpolation weights $w_i$. 


\section{Results}
\label{sect:results}
The dataset that we use to evaluate our B-spline control point interpolation method is generated by a cosmology simulation called $\nu$bhlight \cite{miller2019nubhlight} for solving general relativistic magnetohydrodynamics. The simulation generates particle traces for 2001 time steps. 
We implemented both our method and the baseline method \cite{Chandler2015} with Python for a fair comparison. The B-spline approximation is performed using SciPy API to FITPACK, which is a Fortran routine for B-spline fitting and evaluation. We calculate the knot vectors for each curve before fitting by implementing the method described by fast automatic knot placement\cite{yeh2020fast}. Our implementation can be found here\footnote{\url{https://github.com/harviu/interp\_based\_spline\_tracing}}.

\subsection{B-spline Fitting Evaluation}
\label{sect:fitting_eval}

As the first step of our approach, we evaluate the quality of the B-spline approximation on pathlines. The approximation accuracy depends on the knot placement and how we parameterize the curve. We calculated the fitting error as root-mean-squared-error(RMSE) across different integration steps over time.
Overall, we achieve RMSE of $1.31 \times 10^{-5}$, about $0.000095\%$ of the data range, across all time steps using the 3D B-spline curves with 100 control points parameterized by time.

\begin{figure}
 \centering
 \includegraphics[width=\columnwidth]{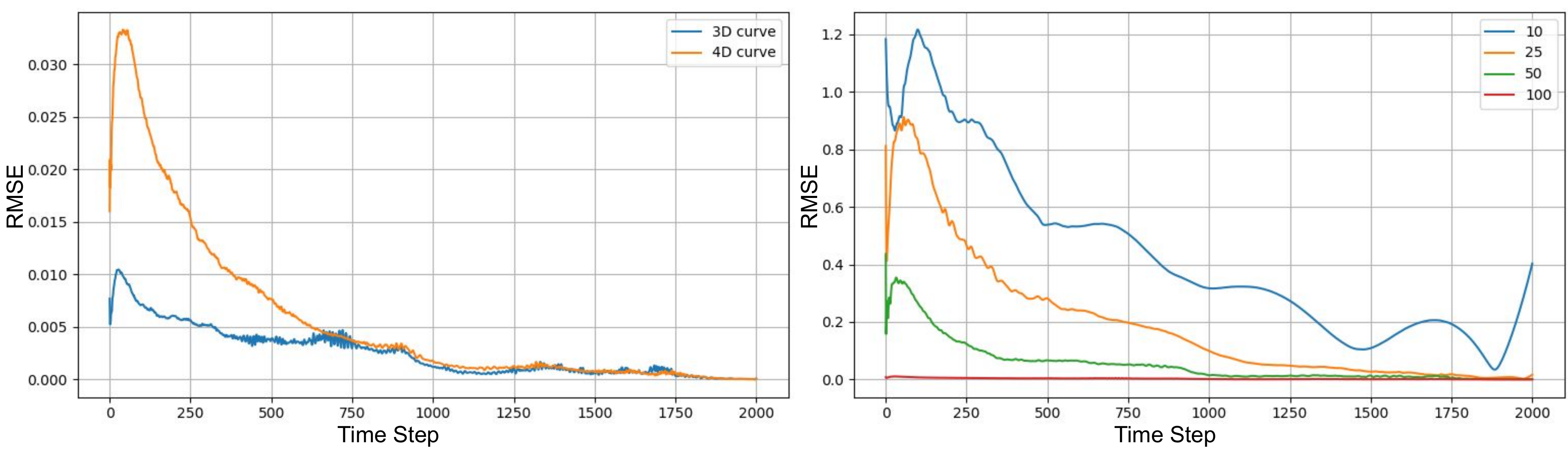}
 \vspace{-20pt}
 \caption{
    B-spline fitting error in RSME. The left figure compares the 3D curve parameterized by time and the 4D curve parameterized by chord length. The right figure compares the 3D curve using different numbers of control points.
 }
 \label{fig:fitting_accuracy}
\end{figure}

We compared the fitting accuracy under two different conditions: 4D curves parameterized by the chord length and 3D curves parameterized by time with 100 control points as shown in the left part of \autoref{fig:fitting_accuracy}, and four different numbers of control points for 3D curves in the right figure. 
\clr{These errors are calculated by interpolating the B-spline curves at the parameters $u_0, u_1, ..., u_{m-1}$, which corresponds to the time steps in B-spline fitting. 
Since 4D errors are not comparable to 3D errors, we only calculate the 3D spatial error in the 4D B-spline case.}
We can observe that 3D curves have fewer errors, especially at the first 750 time steps, when the pathlines have higher curvature in the dataset. The reason for the error difference is that 4D spatial-temporal curves have more complicated geometry because of the additional dimension, which means more control points are needed to achieve a similar spatial error. 
The comparison between the different number of control points clearly shows that increasing from 10 to 100 control points dramatically decreases the approximation error. However, using more than 100 control points does not decrease the error much, while increasing the computational burden in spline tracing. Since the RMSE of the time step with the largest error is already low enough ($0.01$, $0.07\%$ of the data range), we choose to use $100$ control points for the latter experiments. For other datasets or under different use cases, we may also use a linear regression model as a heuristic method \cite{yeh2020fast} to determine the number of control points.

\subsection{Interpolation-based B-spline Tracing Results}

In this section, we quantitatively and qualitatively compare the B-spline tracing results with the particle tracing results. For all $83274$ pathlines in the dataset, we randomly sample $25\%$ of them as the test data. We start the pathline tracing at different time steps using the sampled test data position. 
We use the same inverse distance weighting (IDW) interpolation for both methods for a fair comparison. 
The pathline tracing results are compared with the ground truth test data to calculate spatial RMSE across different time steps.  

\begin{figure}
 \centering
 \includegraphics[width=\columnwidth]{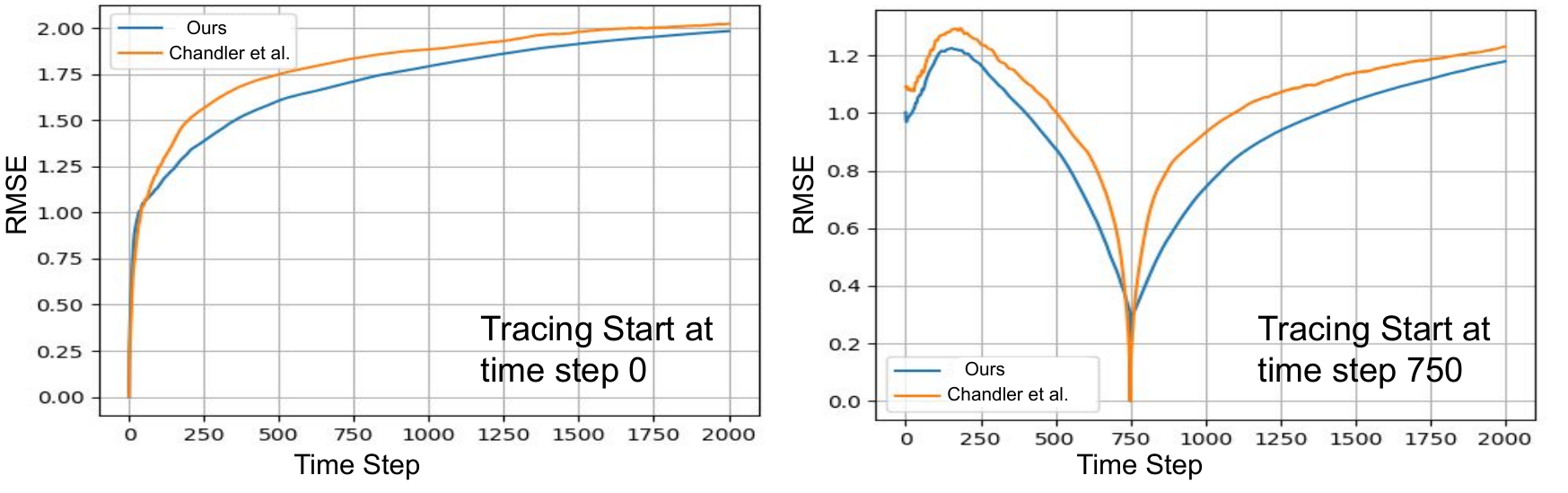} 
 \vspace{-20pt}
 \caption{
    The pathline tracing results that start at different time steps (0 and 750) using our method and the method proposed by Chandler et al. \cite{Chandler2015}. Our method has a slightly lower error under almost all conditions. 
 }
 \label{fig:tracing_error}
\end{figure}

\begin{figure}
 \centering
 \includegraphics[width=\columnwidth]{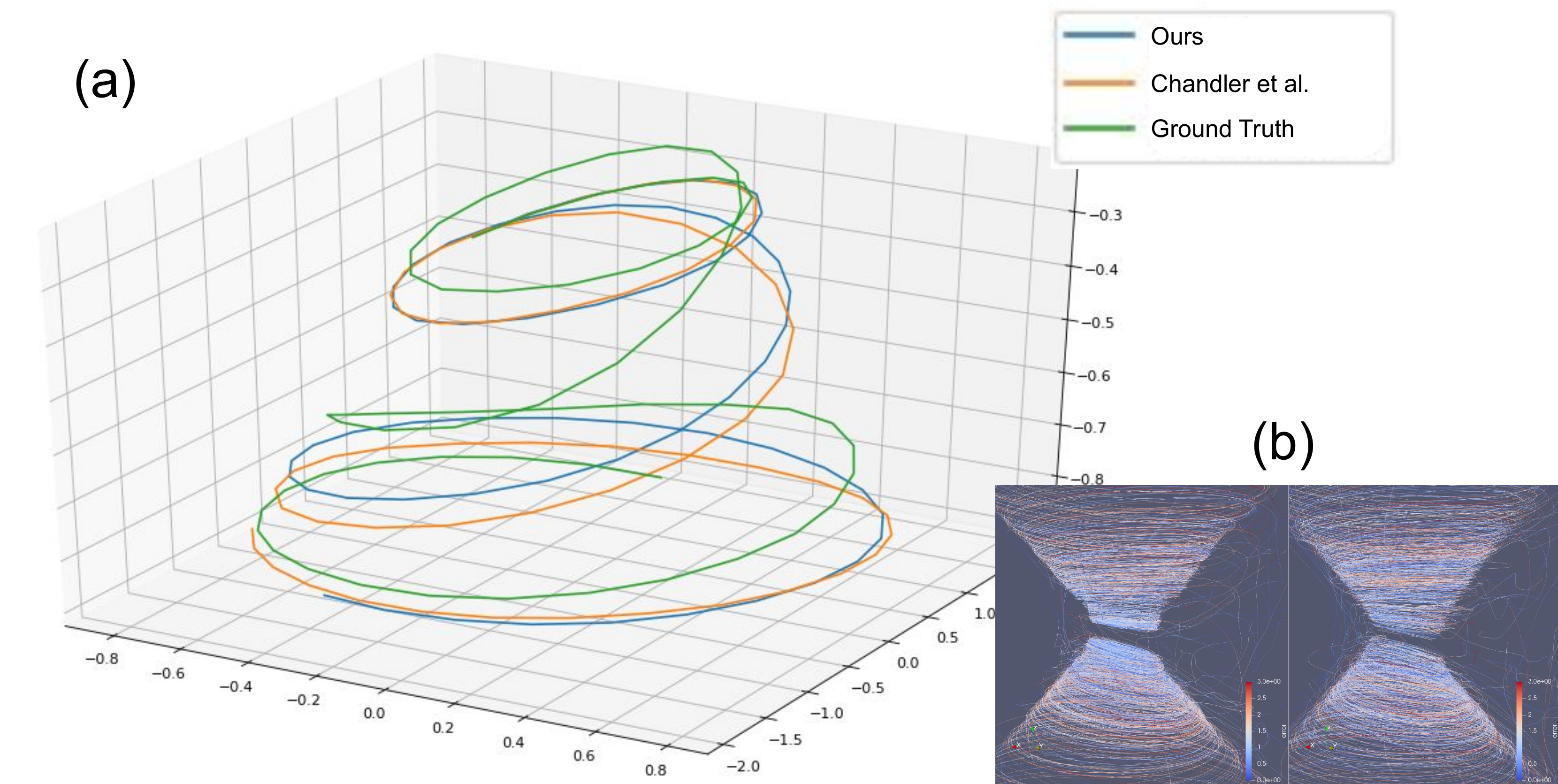} 
 \vspace{-20pt}
 \caption{
The single pathline tracing results (a) and the tested pathline tracing results (b). Our method is on the left and the baseline approach is on the right. Color denotes the error to the ground truth pathlines. Two methods have similar results both in (a) and (b).
 }
 \label{fig:qualitative}
\end{figure}

Quantitative results are shown in \autoref{fig:tracing_error}. Our method and the compared method have a similar error trend when tracing from different starting points. Errors are low around the starting points and increase when the traces advance further. Errors are generally smaller in the first 100 time steps because the data ranges are smaller in these time steps, which is the property of this dataset. 
Our method has slightly lower tracing error across all time steps, possibly because the B-spline is a smoother representation of the original particle trajectories, and thus removing the fluctuations in the original trajectories could lead to lower error. 
Next, we show the qualitative pathline tracing results using two different methods in \autoref{fig:qualitative}. 
\clr{In \autoref{fig:qualitative} (a), we show three pathlines (first 100 time steps) generated by our method, by the baseline method, and from the original test data. The pathline traced using our method and the baseline method are similar. The errors between the traced pathline and the ground truth pathline are accumulated for later time steps. However, this problem exists for both approaches and can be eased by using a more sophisticated interpolation method.
In \autoref{fig:qualitative} (b), we show all the traced pathlines from the test dataset. }
The color denotes the tracing error at different time steps for each pathline. We can observe similar tracing quality using two different methods. 
Since it has already been shown\cite{Chandler2015} that the baseline method has similar tracing accuracy compared to numerical integration methods like adaptive Runge-Kutta 4/5 \cite{cash1990variable}, our method has a comparable tracing error with both the baseline method \cite{Chandler2015} and numerical integration, and moreover, requires much less computation, which we will show in the next section.

\subsection{Timing}
We calculate the B-spline fitting time and the interpolation time of 20818 test pathlines. All the experiments are performed on the same machine and with a similar implementation optimization. The particle dataset is assumed to be pre-arranged in a way that particles in the same pathline are close to each other in the memory. 

We present the B-spline control points tracing computational time under the conditions of different numbers of control points in \autoref{tab:time}. 
The number of control points does not influence the B-spline fitting time much. However, the interpolation time increases linearly with the number of control points of interpolation. 
For the number of control points 100 that we choose for the experiments, the interpolation time is about 10\% of the particle interpolation time. 
\clr{For most flow datasets, choosing a number of control points similar to this dataset (~5\% of the number of time steps) can achieve high B-spline fitting accuracy \cite{bujack2015lagrangian}. If the flow data is especially complex, e.g. a lot of fluctuation along the particle trajectories, and we can not reduce the number of control points much, the computation cost of our method could exceed the computation cost of directly interpolating particles. However, this is unlikely for real flow simulations. 
The B-spline fitting time is much less than that of interpolation and the fitting only needs to be done once for the same dataset.
In terms of scalability, the time for the fitting process and tracing for different pathlines increases linearly to the number of pathlines. Since the neighborhood search takes most of the time for interpolating between control points, the time will increase logistically to the number of pathlines.
}



\section{Conclusion and Future Works}

We presented an interpolation-based pathline tracing method for particle simulations on B-spline control points. 
The B-spline approximation is optimized by using an adaptive knot placement method, which is fast and has controllable errors. The computation time for our method is largely reduced due to the reduction of interpolation iteration, and at the same time, our method achieves similar pathline tracing quality compared to interpolating particles.
Besides the usage for pathline tracing, B-splines can also be applied to data reduction or as a proxy representation for neural network training. 
Our future works will focus on B-spline and neural network representations for the pathlines and their application for feature-driven visualization.

\begin{table}[tb]
  \caption{
  B-spline approximation time and control points interpolation time under the conditions of different numbers of control points. 
  The ratio shows the computation time compared to the baseline method.
  }
  \label{tab:time}
  \scriptsize
	\centering
  \begin{tabularx}{\columnwidth}{cccc}
  \toprule
Number of Control Points  & Fitting Time  & Interpolation Time & Ratio \\
  \midrule 
    10 &  67.36s & 3.19s & 0.011 \\
    25 &  67.52s & 6.50s & 0.023 \\ 
    50 &  68.80s & 11.04s & 0.040 \\
    100 & 69.28s & 26.71s & 0.096 \\
    Baseline\cite{Chandler2015} & -  & 278.90s & 1 \\
  \bottomrule
  \end{tabularx}
\end{table}
\acknowledgments{
This work is supported in part by US Department of Energy SciDAC program DE-SC0021360, National Science Foundation Division of Information and Intelligent Systems IIS-1955764, and National Science Foundation Office of Advanced Cyberinfrastructure OAC-2112606.
}

\bibliographystyle{abbrv-doi}

\bibliography{template}
\end{document}